\documentclass[fleqn,12pt]{article}
\usepackage[utf8]{inputenc}
\usepackage[left=18mm, top=18mm, right=15mm, bottom=20mm,]{geometry}
\usepackage{amsmath,amssymb}
\usepackage{esint}
\usepackage{amsfonts}
\usepackage{graphicx}
\usepackage{cite}
\usepackage{hyperref}
\usepackage{xcolor}
\usepackage{caption}
\definecolor{linkcolor}{named}{black}
\hypersetup{citecolor=black, colorlinks=true, linkcolor=black}
\usepackage{setspace}
\usepackage{ragged2e}
\justifying          
\newcommand{\bea}{\bear     }
\newcommand{\eea}{\ear     }

\usepackage{amsfonts,amssymb,cite}
\usepackage{graphicx}

\def\onecol{\onecolumn \mathindent 2em}

\def\noi{\noindent}

\newcommand{\Title}[1]{\noi {{\Large\bf #1}}\\[1ex]}

\def\Aunames#1{\noi{\bf #1}}

\def\Addresses#1{\medskip\noi \protect
	\begin{description}\itemsep -3pt {\it #1} \end{description}}
\def\adr#1#2{\item[${}^{#1}$]{\it #2}}

\newcommand{\Abstract}[1]{\vskip 2mm \begin{center}
        \parbox{16.4cm}{\small\noi #1} \end{center}\medskip}

\def\email#1#2{\footnotetext[#1]{e-mail: #2}\addtocounter{footnote}{1}}

\def\nqq{\hspace*{-2em}}

\def\qq{\qquad}

\def\deg{\mbox{${}^\circ$}}                     
\usepackage{color}

\def\Jl#1#2{#1 {\bf #2},\ }

\def\ApJ#1 {\Jl{Astroph. J.}{#1}}
\def\CQG#1 {\Jl{Class. Quantum Grav.}{#1}}
\def\DAN#1 {\Jl{Dokl. AN SSSR}{#1}}
\def\GC#1 {\Jl{Grav. Cosmol.}{#1}}
\def\GRG#1 {\Jl{Gen. Rel. Grav.}{#1}}
\def\IJMPD#1 {\Jl{Int. J. Mod. Phys. D}{#1}}
\def\JETF#1 {\Jl{Zh. Eksp. Teor. Fiz.}{#1}}
\def\JETP#1 {\Jl{Sov. Phys. JETP}{#1}}
\def\JHEP#1 {\Jl{JHEP}{#1}}
\def\JMP#1 {\Jl{J. Math. Phys.}{#1}}
\def\NPB#1 {\Jl{Nucl. Phys. B}{#1}}
\def\NP#1 {\Jl{Nucl. Phys.}{#1}}
\def\PLA#1 {\Jl{Phys. Lett. A}{#1}}
\def\PLB#1 {\Jl{Phys. Lett. B}{#1}}
\def\PRD#1 {\Jl{Phys. Rev. D}{#1}}
\def\PRL#1 {\Jl{Phys. Rev. Lett.}{#1}}

\def\lal{&&\nqq {}}
\def\eq{Eq.\,}
\def\eqs{Eqs.\,}
\def\beq{\begin{equation}}
\def\eeq{\end{equation}}
\def\bear{\begin{eqnarray}}
\def\bearr{\begin{eqnarray} \lal}
\def\ear{\end{eqnarray}}
\def\earn{\nonumber \end{eqnarray}}

\def\nnn{\nonumber\\ \lal }

\def\yy{\\[5pt] {}}

\def\const{{\rm const}}

\def\then{\ \Rightarrow\ }

\def\rf{\eqref}

\def\sph{spherically symmetric}

\def\wh{wormhole}
\def\whs{wormholes}

\def\emag{electromagnetic}

\def\cosm{cosmological}

\def\Lem{Lema\^{\i}tre}

\begin{document}
\thispagestyle{empty}
\onecol
\vspace{-3mm}

\Title{Gravitational lensing by \Lem-Tolman-Bondi wormholes \yy
	in a Friedmann universe
}

\Aunames{Kirill A. Bronnikov,$^{a,b,c,1}$ Valeria A. Ishkaeva,$^{d,2}$ Sergey V. Sushkov$^{d,3}$}

\Addresses{\small
	\adr a {Center of Gravitation and Fundamental Metrology, Rostest, 
			Ozyornaya ul.~46, Moscow 119361, Russia}
	\adr b {Institute of Gravitation and Cosmology, Peoples' Friendship University of 
		Russia (RUDN University), ul. Miklukho-Maklaya~6, Moscow 117198, Russia}
	\adr c  {National Research Nuclear University ``MEPhI'',  
			Kashirskoe Shosse~31, Moscow 115409, Russia}
	\adr d  {Institute of Physics, Kazan Federal University, 
			Kremliovskaya~ul.~16a, Kazan 420008, Russia}
		}

\Abstract
{The \Lem-Tolman-Bondi (LTB) solution to the Einstein equations describes the dynamics of a self-gravitating \sph\ dust cloud with an arbitrary density profile and any distribution of initial velocities, encoded in three arbitrary functions 
$f(R)$, $F(R)$, and $\tau_0(R)$, where $R$ is a radial coordinate in the comoving 
reference frame.
A particular choice of these functions corresponds to a wormhole geometry with a throat defined as a sphere of minimum radius at a fixed time instant. 
In this paper we explore LTB wormholes and discuss their possible observable appearance studying in detail the effects of gravitational lensing by such objects.
For this aim, we study photon motion in wormhole space-time inscribed in a closed Friedmann dust-filled universe and find the wormhole shadow as it could be seen 
by a distant observer.
Since the LTB wormhole is a dynamic object, we analyze the dependence of its shadow size on the observation time and on the initial size of the wormhole region.  
We reveal that the angular size of the shadow exhibits a non-monotonic dependence on the observation time. At early times, the shadow size decreases as photons with smaller angular momentum gradually reach the observer. At later times, the expansion of the Friedmann Universe becomes a dominant factor that leads to an increase in the shadow size.
}

\email 1 {kb20@yandex.ru}
\email 2 {ishkaeva.valeria@mail.ru}
\email 3 {sergey\_sushkov@mail.ru}

\section{Introduction}

  Wormholes are hypothetical astrophysical objects resembling tunnels that connect two 
  different regions of the same space-time or two different space-times. A basic problem 
  of wormhole physics, at least in the case of static configurations in general relativity
  (GR), is that a wormhole needs exotic matter which violates the null energy condition 
  (NEC) and prevents the throat from collapsing \cite{MorrisThorne1, hoh-vis1}. 

  There have been numerous attempts to circumvent the theorems on NEC violation 
  either by invoking extensions of GR (which are desirable for many reasons but are quite
  unnecessary as regards the empirical situation on the macroscopic level \cite{macro})
  or by abandoning the assumption on the static nature of a \wh. A minimal way to do that 
  is to consider stationary systems invoking spin or rotation, and indeed, such examples 
  of \wh\ solutions in GR have been obtained, in particular, those with classical spinor fields 
  \cite{Radu21, KonZh21, we21} and with rotating cylindrical sources \cite{Kr1, Kr2}. It may
  be remarked, however, that such matter sources without exotic matter, being of evident 
  theoretical interest, still look rather unrealistic from an observational viewpoint.     
 
  It is evidently more promising to obtain \wh\ models without exotic matter by considering 
  manifestly dynamic systems. In such cases, \whs\ in the framework of GR cannot exist 
  eternally \cite{hoh-vis2} but their lifetime may be sufficiently long from any practical
  viewpoint. Thus, examples of \whs\ existing against a cosmological background, sourced
  by some special examples of nonlinear \emag\ fields, were found in 
  \cite{ArLobo06, kb18}. It has also turned out that dynamic wormhole configurations
  can even be obtained with such a familiar source of gravity as evolving dust clouds,   
  as follows from the recent papers  \cite{KasSus, BroKasSus:2021,BroKasSus:2023}.
  Such models were found there as particular cases of the \Lem-Tolman-Bondi (LTB) solution, 
  obtained in GR by {\Lem} and Tolman in 1933-1934 
  \cite{Lemaitre:1933gd, Tolman:1934za} and studied by Bondi in 1947 \cite{Bondi:1947fta},
  as well as its extensions to a nonzero \cosm\ constant and an \emag\ field.
  (One more recent paper \cite{ChaCha} also claimed to study LTB \whs\ but actually 
  considered mixtures of fluids with nonzero pressure in a \cosm\ background.)

  The LTB solution describes the evolution of a spherical dust cloud. It contains three arbitrary 
  functions $f(R)$, $F(R)$, and $\tau_0(R)$, where $R$ is a radial coordinate in the comoving 
  reference frame.  A particular choice of these functions corresponds to a wormhole
  geometry with a throat defined as a sphere of minimum radius at a fixed time instant. 

  The normal vector to a throat of a dynamic \wh\ is timelike, hence a throat is in general
  located in a T-region of space-time. Thus if such a dust cloud is placed between two 
  empty Schwarzschild space-time regions, the whole configuration is a black hole rather
  than a wormhole. However, dust  clouds with throats can be inscribed into closed 
  isotropic cosmological models filled with dust to form wormholes which exist for a finite
  period of time and experience expansion and contraction 
  together with the corresponding cosmology. 

  In Ref. \cite{BroKasSus:2023}, we studied in detail evolving wormholes able to exist in 
  a closed Friedmann dust-filled universe. In particular, we have shown that the lifetime 
  of \wh\ throats is much shorter than that of the whole \wh\ region in the universe 
  (which coincides with the lifetime of the universe as a whole). Nevertheless, studying 
  radial null geodesics, i.e., radial photons paths, we established the possible 
  traversability of the LTB wormhole configurations.  

  In this paper, we continue exploring the LTB wormholes and now discuss their 
  possible observable appearance studying in detail the effects of gravitational
  lensing by such objects.   

  One can note that gravitational lensing by \whs\ is rather widely 
  discussed in the literature, but mostly for static or stationary \wh\ models, see, e..g., 
  \cite{lens1,lens2,lens3,lens4,lens5} and references therein. The problem under consideration 
  here is much more complicated due to the essentially dynamic nature of the \whs.  
  
  The paper is organized as follows. In Section \ref{sec2}, we briefly describe the class of 
  solutions under study. Section 3 is devoted to an analysis of pohoton motion in 
  \wh\ space-time. In Section \ref{sec4}, we describe the procedure of inscribing LTB
  wormholes into a closed Friedmann dust-filled universe and find their shadows as 
  they could be seen by distant observers. Section \ref{sec5} is a conclusion.

\section{The \Lem-Tolman-Bondi solution and wormholes \label{sec2}}

  Consider the \Lem-Tolman-Bondi (LTB) solution to the Einstein equations 
  \cite{Tolman:1934za, Lemaitre:1933gd, Bondi:1947fta}, 
  describing the dynamics of a \sph\ dust cloud with an arbitrary density profile and 
  any distribution of initial velocities. In a reference frame comoving to the dust particles,
  the metric can be written in the form \cite{Landau:1975pou}:
\beq      \label{LTBmetric}
	ds^2=d\tau^2-e^{2\lambda(R,\tau)}dR^2-r^2(R,\tau)(d\theta^2+\sin^2\theta d\phi^2),
\eeq      
  where $\tau$ is physical time measured by clocks attached to dust particles, while $R$ is
  the radial coordinate whose values correspond to fixed Lagrangian spheres, i.e., 
  such a sphere of particles does not change its particular value of $R$ during its motion. 
  
  We are here interested in dynamic \whs\ described by the metric \rf{LTBmetric} which, 
  according to \cite{BroKasSus:2021,BroKasSus:2023}, are only possible in the elliptic 
  branch of the LTB solution. Therefore we here consider only this branch, for which 
  (assuming the \cosm\ constant $\Lambda=0$) we can write the 
  following first integrals of the Einstein equations \cite{BroKasSus:2023}:
\bear     \label{drdR} 
	e^{2\lambda(R,\tau)}&=&\frac{r'^2(R,\tau)}{1-h(R)},
\\	        \label{drdt}
	\dot{r}^2(R,\tau)&=&\frac{F(R)}{r(R,\tau)}-h(R),
\ear     
  where $r'\equiv\partial r/\partial R$, $\dot{r}\equiv\partial r/\partial\tau$, $F(R)$ and 
  $h(R)$ are arbitrary functions, such that $F(R)$ is responsible for the mass distribution,
  and $0<h(R)<1$ for the distribution of initial velocities. 
  
  Integrating \eq (\ref{drdt}), we obtain an implicit expression for the radius $r(R, \tau)$
\beq      \label{tr}
	\pm[\tau-\tau_0(R)]=\frac{1}{h}\sqrt{Fr-hr^2}+\frac{F}{2h^{3/2}}\arcsin\frac{F-2hr}{F},
\eeq      
  where  $\tau_0(R)$ is one more arbitrary function, responsible for clock synchronization 
  between different Lagrangian spheres. 
  It is more convenient to rewrite this solution in a parametric form in terms of the parameter 
  $\eta$ \cite{Landau:1975pou}:
\bearr      
		r=\frac{F}{2h}(1-\cos\eta),
\nnn            \label{rtn}
		 \pm[\tau-\tau_0(R)]=\frac{F}{2h^{3/2}}(\eta-\sin\eta).
\ear

   In wormhole solutions the arbitrary functions must satisfy certain conditions. Thus, 
   at a throat,  defined as a minimum of $r$ at fixed $\tau$,  we must have 
   \cite{BroKasSus:2021,BroKasSus:2023} 
\bearr                    \label{throat}
		h = 1, \quad\ h'=0, \quad\ h'' < 0,
\nnn		
		F' = 0, \quad\ r' =0,\quad\   \frac{h'}{r'} < 0, \quad\  \frac{F'}{r'} > 0. 
\ear
  It is also supposed that the size $r$ of a fixed-time section of space-time is much larger 
  on both sides of the throat than the size $r\big|_{\rm th}$ of the throat itself.

  In what follows we will solve geodesic equations in the \wh\ branch of the solution under 
  study in the approximation of small $\eta$,  thus considering only the beginning of 
  \wh\ evolution,  and assuming also $\tau_0(R)=0$.
  Then \eqs (\ref{rtn}) are rewritten as
\beq      \label{rtnapp}
	\tau\approx\frac{F\eta^3}{12h^{3/2}},\qq
	r\approx \frac{F \eta^2}{4h}               \ \ \then \ \ 
	r\approx \Big(\frac 32\Big)^{2/3}F^{1/3}\tau^{2/3}.
\eeq      

  The next terms in the $\eta$ expansions of $\tau$ and $r$ are, respectively,  
\[
	-\dfrac{F\eta^5}{240\,h^{3/2}}  \ \ \ {\rm and} \ \ \  - \dfrac{F\eta^4}{48 h}.
\]      
   Their ratios to the first terms \rf{rtnapp} may be considered as relative errors of our
   approximation at given $\eta$. We will construct photon trajectories for $\eta \leq 0.1$, 
   so the relative errors in the quantities $\tau$ and $r$ will not exceed $1/2000$ and
   $1/1200$, respectively. Thus we can hope that the results of all further calculations 
   will be correct up to $\sim 10^{-3}$.

\section{Photon motion in \wh\ space-times \label{sec3}}
\subsection{Null geodesic equations} 
  
   The geodesic equations $\frac{d^2x^i}{ds^2}+\Gamma^i_{k l}\frac{dx^k}{ds}\frac{dx^l}{ds}=0$,
   where $\Gamma^i_{k l}$ are Christoffel symbols and $s$ the affine parameter, 
   have the following form for the metric (\ref{LTBmetric}):
\bear     \label{dt}
	\frac{d^2\tau}{ds^2}&=&-\dot{\lambda}e^{2\lambda}\left(\frac{dR}{ds}\right)^2-r\dot{r}
	\left[\left(\frac{d\theta}{ds}\right)^2+\sin^2\theta \left(\frac{d\phi}{ds}\right)^2\right],
\\          \label{dR}
	\frac{d^2R}{ds^2}&=&-2\dot{\lambda}\frac{d\tau}{ds}\frac{dR}{ds}
	-\lambda'\left(\frac{dR}{ds}\right)^2+rr'e^{-2\lambda}\left[\left(\frac{d\theta}{ds}\right)^2
	+\sin^2\theta \left(\frac{d\phi}{ds}\right)^2\right],
\\	
	\frac{d^2\theta}{ds^2}&=&-2\frac{\dot{r}}{r}\frac{d\theta}{ds}\frac{d\tau}{ds}
	-2\frac{r'}{r}\frac{d\theta}{ds}\frac{dR}{ds}+\sin\theta\cos\theta\left(\frac{d\phi}{ds}\right)^2,
\\
	\frac{d^2\phi}{ds^2}&=&-2\frac{\dot{r}}{r}\frac{d\phi}{ds}\frac{d\tau}{ds}
	-2\frac{r'}{r}\frac{d\phi}{ds}\frac{dR}{ds}-2\cot\theta\frac{d\theta}{ds}\frac{d\phi}{ds}.
\ear     

   Due to spherical symmetry, as usual, without loss of generality we can consider particle 
   motion in 
   the equatorial plane $\theta=\pi/2$. Let us also try to reduce the system to first-order 
   differential equations. Since $\phi$ is a cyclic coordinate, there is an integral of motion of 
   the form $p_{\phi}=r^2\frac{d\phi}{ds}=\const=L$, where $L$ is the asimuthal angular 
   momentum of a particle.
   
   Let us write the Lagrangian for photon motion:
\beq      \label{L}
	2\mathcal{L}=\left(\frac{d\tau}{ds}\right)^2
	-e^{2\lambda}\left(\frac{dR}{ds}\right)^2-r^2\left(\frac{d\phi}{ds}\right)^2=0.
\eeq   
   Its derivative in the affine parameter $s$ is actually a combination of \eqs (\ref{dt}) and (\ref{dR}), 
   hence we can replace one of them with (\ref{L}). Then, with \eq (\ref{drdR}), the geodesic 
   equations for photons moving in the equatorial plane can be rewritten as
\bear       \label{dt2}
	\frac{d^2\tau}{ds^2}&=&-\frac{\dot{r}'}{r'}\left(\frac{d\tau}{ds}\right)^2
	+\frac{L^2}{r^2}\left(\frac{\dot{r}'}{r'}-\frac{\dot{r}}{r}\right),
\\          \label{dR2}
	\frac{dR}{ds}&=&\pm\sqrt{\frac{1-h}{r'^2}\left[\left(\frac{d\tau}{ds}\right)^2
	-\frac{L^2}{r^2}\right]},
\\           \label{df2}
	\frac{d\phi}{ds}&=&\frac{L}{r^2}.
\ear      

\subsection{Geodesics in the small $\eta$ approximation}

  Let us begin with \eq (\ref{dt2}) and notice that ${\dot{r}'}/{r'}={\dot{r}}/{r}={2}/(3\tau)$
  (see Eq. (\ref{rtnapp})), i.e., in our approximation $L$ disappears from 
  \eq \rf{dt2} that now takes the form
\beq      
		\frac{d^2\tau}{ds^2}=-\frac{2}{3\tau}\left(\frac{d\tau}{ds}\right)^2.
\eeq      
  This equation is easily integrated giving
\beq      \label{tau}
		\tau(s)=\left(C\,s+s_0\right)^{3/5},
\eeq      
  where $s_0$ is related to the photon launching time, 
  $s_0=\tau(0)^{5/3}$, and $C$ is related to the initial photon energy $E_0$, such that 
  $C=(5/3)\tau(0)^{2/3}(d\tau/ds)|_{s=0}=(5/3)\tau(0)^{2/3}p_{\tau}(0)=(5/3)\tau(0)^{2/3}E_0$.
    
   Let us now address \eq (\ref{dR2}). Taking into account (\ref{tau}) and substituting the 
   expression (\ref{rtnapp}) for $r$, we can rewrite \eq (\ref{dR2}) as
\beq \label{dR/ds}      
	\frac{dR}{ds}=\pm\frac{C\,(18\,F)^{2/3}}{5\,|F'|}
	(1-h)^{1/2}
	\sqrt{1-\frac{L^2}{C^2}\left(\frac{500}{243\,F}\right)^{2/3}}\tau^{-4/3}.
\eeq      
  It is important to notice here that the right-hand side in (\ref{dR/ds}) can turn to zero if the 
  expression under the square root vanishes. It is possible at a point $R=R_t$ such that 
\beq      
             F(R_t)=\frac{500}{243}\,\left|\frac{L}{C}\right|^3.
\eeq 
  The condition $dR/ds|_{R=R_t}=0$ means that a photon trajectory with given $L$ has 
  a turning point. If one supposes that the photon moves initially inward, i.e. from spheres 
  with larger to smaller radial coordinates $R$, then the turning point is a sphere with a
  minimum coordinate $R_t$, where the inward motion of the photon stops and changes 
  to the outward one.  

  Integrating \eq \rf{dR/ds}, we obtain
\beq      \label{dR4}
	\pm\fint_{R(0)}^{R}\frac{|F'|\,d\tilde{R}}{12^{1/3}F^{2/3}\sqrt{(1-h)
	\left[1-\dfrac{L^2}{C^2}\left(\dfrac{500}{243\,F}\right)^{2/3}\right]}}=3\left(C\,s+s_0\right)^{1/5},
\eeq      
  where $\fint_{R(0)}^{R}=\int_{R(0)}^{R_t}+\int_{R_t}^{R}$.

\subsection{Photon paths in dynamic \wh\ space-time}

   Thus far we did not specify the functions $F(R)$ and $h(R)$. Now, following 
   Ref.\,\cite{BroKasSus:2023}, we choose them as follows:
\beq                             \label{WH}
	F=2b(1+R^2)^k,\qq          h=\frac{1}{1+R^2},
\eeq      
  with  $b,k = \const >0$. 
  The function $h(R)$ is taken in this form without loss of
  generality due to arbitrariness of $R$ parametrization (but according to the \wh\
  existence conditions \rf{throat}), whereas the choice of $F(R)$ is significant: the parameter 
  $b$ is the maximum size of the throat while $k$ is responsible for the \wh\ density. 
  Then the function $r(R,\tau)$ in the small $\eta$ approximation is
\beq      
	r(R,\tau)\approx (4.5\,b)^{1/3}\left(1+R^2\right)^{k/3}\tau^{2/3}.
\eeq      
  Under this choice, the metric  (\ref{LTBmetric}) describes the space-time of a dynamic \wh,
  is symmetric with respect to its throat $R = 0$. The turning point of a photon path with specified 
  $L$ and $C$ is given by 
\beq      \label{Rturn}
	 R_t=\pm\sqrt{\left(\frac{250}{243\,b}\right)^{1/k}\left|\frac{L}{C}\right|^{3/k}-1}.
\eeq      

\begin{figure}[h!]
\centering
\includegraphics[width=0.99\linewidth]{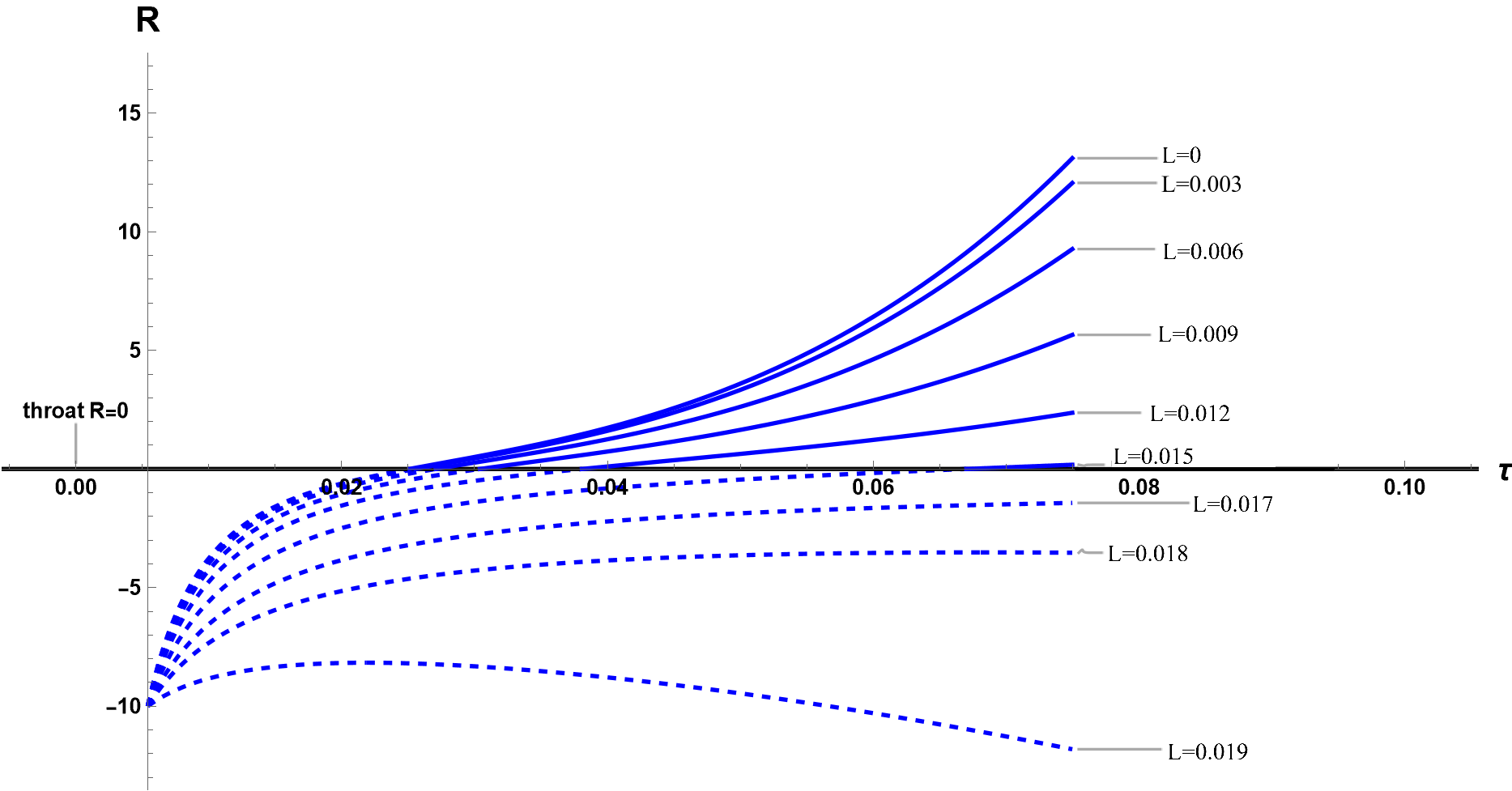}
\caption{\small
		Paths $R(\tau)$ of photons with different $L$. Dashed lines mark the photon motion at 
		$R<0$. The \wh\ parameters are $b=1, k=0.1$. The photons are launched at $R_0=-10$,
		$\tau_0\approx 0.005$. All photons have the same initial energy $E_0\approx0.3$.}
	\label{Rt10traj}
\end{figure}

\begin{figure}[h!]
\centering
\includegraphics[width=0.99\linewidth]{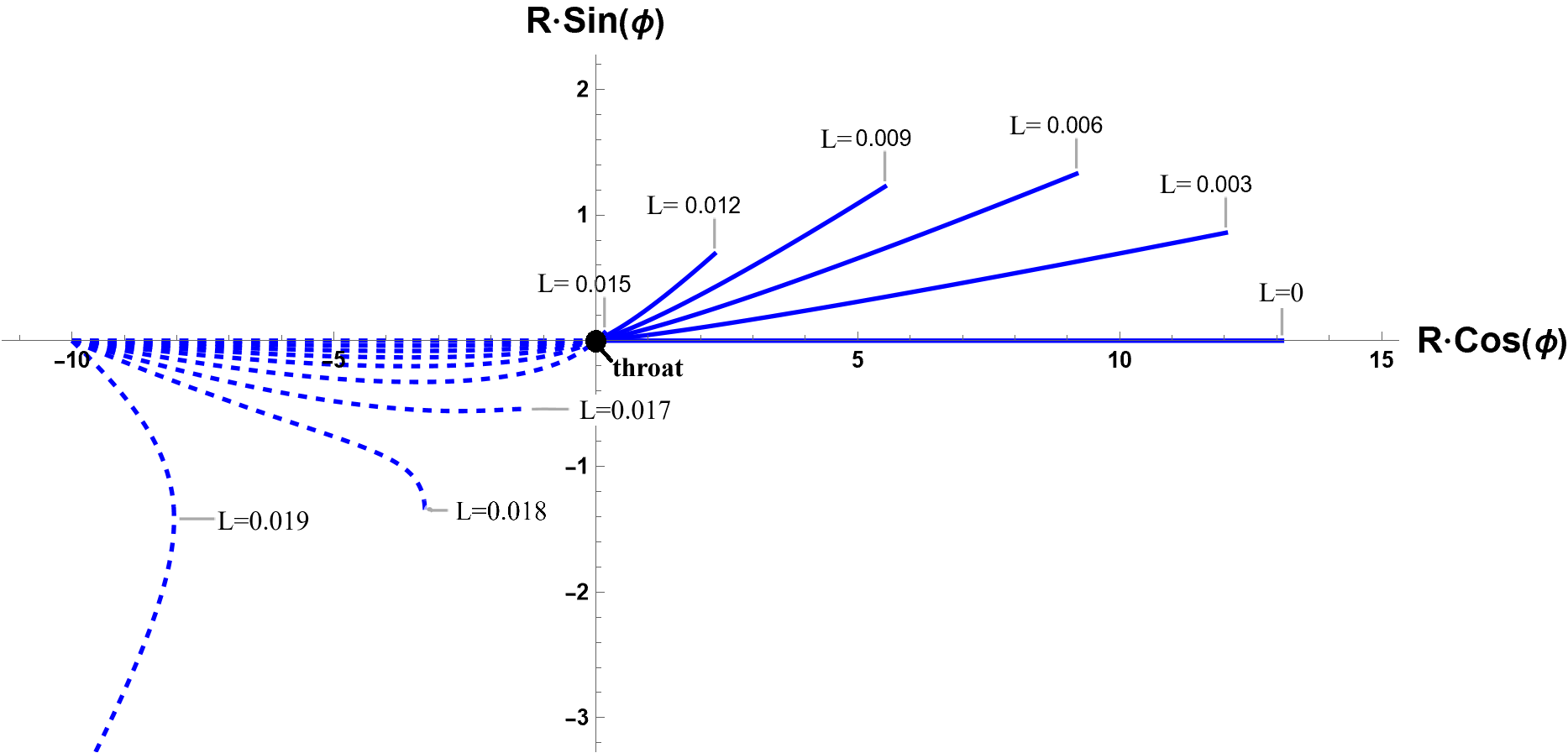}
\caption{\small
		Paths $R(\phi)$ of photons with different $L$. Dashed lines mark the photon motion at 
		$R<0$. The \wh\ parameters are $b=1, k=0.1$. The photons are launched at $R_0=-10$,
		$\tau_0\approx 0.005$. All photons have the same initial energy $E_0\approx0.3$.}
\label{Rf10traj}
\end{figure}

  Figures  \ref{Rt10traj} and \ref{Rf10traj} present examples of photon paths $R(\tau)$ and $R(\phi)$ 
  in a \wh\ with the parameters  $b=1, k=0.1$. Dashed lines mark the photon motion in the region 
  $R<0$. All photons are launched at $R_0=-10,\ \tau_0\approx 0.005$ with the initial energy  
  $E_0\approx0.3\ (C\approx0.017)$. With such parameters, photons with $|L|>0.0165$ have a 
  turning point, hence they cannot cross the throat and remain in the region $R<0$.
\\

\section{A Friedmann universe with a dynamic \wh \label{sec4}}

  In what follows we are going to study photon paths in a Friedmann universe containing a dynamic 
  \wh, therefore, let us first consider null geodesics in the Friedmann metric.

  The metric of a closed isotropic Friedmann universe filled with dustlike matter can be obtained 
  from the LTB metric (\ref{LTBmetric}) by choosing the arbitrary functions $h$ and $F$ as
\beq          \label{F,h-Fr}
             F(\chi)=2a_0\sin^3\chi, \quad h(\chi)=\sin^2\chi, \quad a_0=\const,
\eeq      
  where the coordinate $R\equiv\chi$ is the radial angle. Then, in terms of the parameter $\eta$, 
  assuming $\tau_0(R)=0$ in \eq (\ref{tr}), we can write the functions $r$ and $\tau$ in the form 
\bearr      
		r(\eta,\chi)=a(\eta)\sin\chi, 
\nnn 
		a(\eta)=a_0(1-\cos\eta), 
\nnn 
		\tau(\eta)=a_0(\eta-\sin\eta),
\ear      
  where $a(\eta)$ is the \cosm\ scale factor, and the Friedmann metric can be written as
\beq      \label{dsFU}
	ds^2=a^2(\eta)\left[d\eta^2-d\chi^2-\sin^2\chi\left(d\theta^2+\sin^2\theta d\phi^2\right)\right].
\eeq      

\subsection{Null geodesics in a Friedmann universe }

  The geodesic equations for photons in a Friedmann universe can be obtained by substituting the
  functions $r(\eta,\chi)$ and $\tau(\eta)$ into the geodesic equations (\ref{dt2})--(\ref{df2}) for 
  the LTB metric. Then, we  can integrate the equation for $\tau$ and write the geodesic equations 
  as follows:
\bea
	\frac{d\eta}{ds}&=&\frac{\sqrt{K}}{a^2(\eta)},\label{Eta}
\\
	\frac{d\chi}{ds}&=&\pm\frac{1}{a^2(\eta)}\sqrt{K-\frac{L^2}{\sin^2\chi}}, \label{Chi}
\\
	\frac{d\phi}{ds}&=&\frac{L}{a^2(\eta)\sin^2\chi}=\frac{L}{r^2}.\label{Phi}
\eea
  The constant $K$ in these equations is related to the initial photon energy $E_0$, such that
  $\sqrt{K}=a_0[1-\cos\eta(0)]E_0$, and $L$ is the asimuthal angular momentum of a photon. 
  Next, when constructing images, we will assume $K=1$.

  Note that in \eq (\ref{Chi}), the root expression should not be negative. Therefore, particles with 
  a nonzero angular momentum have a turning point $\chi_t$:
\bea
		\chi_t=\arcsin\left(\sqrt{\frac{L^2}{K}}\right).
\eea

\subsection{A \wh\ in a Friedmann universe }

  When matching the dynamic wormhole solution with that for a Friedman universe, on the boundary
  determined by some values $\chi_*$ and $R_*>0$ of the radial coordinates, the functions $F(R_*)$ 
  and $h(R_*)$ in the wormhole must coincide with $F(\chi_*)$ and $h(\chi_*)$ in the Friedman 
  universe \cite{BroKasSus:2023},
\beq      
		h_*=\sin^2\chi_*,\quad F_*=2a_0\sin^3\chi_*.
\eeq      
  With our choice of the arbitrary functions $F(R)$ and $h(R)$ in (\ref{WH}), these conditions lead 
  to the equalities
\beq      
		R_*=\cot\chi_*,\quad b=a_0(\sin\chi_*)^{3+2k}.                         \label{match}
\eeq      

  The size and other parameters of a \wh\  in a Friedmann universe are estimated in Table 1. Here and
  henceforth, we assume $a_0\sim 10^{28}$ cm, which corresponds to the size of the observable universe.

\begin{table}[h!]
	\label{tabular:timesandtenses}
	\centering
	\caption{\small
		Estimates of the radius $r_*$ of the wormhole region, the boundary values $\chi_*$ 
		and $R_*$ of the radial coordinates, and the angular size of the shadow $d_{\rm sh}$ 
		(for an observer at a point with $\chi_{\rm obs}=1,\,\eta_{\rm obs}=1.1$), for various 
		throat radii $r_{\rm th}$ in the case $k = 0.1$.
	}
	\begin{tabular}{|c c c c c|} 
		\hline
		$r_{\rm th}\big|_{\eta=\pi}=2b$ & $r_*\big|_{\eta=\pi}$ & $\chi_*$\qq & $R_*$\qq 
					& $d_{\rm sh}$\\ 
		\hline\hline
		1 km & $10^{21}$ cm = 338 pc &\ \ $5.2\times10^{-8}$\ \  &\ \  $1.9\times10^7$\ \ 
				& $0.004''$ \\
		10 km & 700 pc & $1.1\times10^{-7}$ & $9.3\times10^6$ & $0.009''$ \\
		$6.4\times 10^{3}$ km & 5.1 Kpc & $8.1\times10^{-7}$ & $1.2\times10^6$ & $0.08''$ \\
		$2.3\times 10^{5}$ km & 16 Kpc & $2.5\times10^{-6}$ & $4\times10^5$ & $0.3''$ \\  
		$695\times 10^{3}$ km & 23 Kpc & $3.5\times10^{-6}$ & $2.9\times10^5$ & $0.4''$ \\
		$10^{7}$ km & 52 Kpc & $8.1\times10^{-6}$ & $1.2\times10^5$ & $1''$ \\
		$7\times 10^{7}$ km & 96 Kpc & $1.5\times10^{-5}$ & $6.8\times10^4$ & $1.8''$ \\
		1 pc & 5.7 Mpc & $8.6\times10^{-4}$ & 1165 & $2.3'$ \\
		6.5 pc & 10 Mpc & 0.0015 & 648.8 & $4.3'$ \\
		10 Kpc & 100 Mpc & 0.015 & 65.5 & $0.8\deg$ \\
		\hline
	\end{tabular}
	\label{estimates}
\end{table}

\subsection{Shadow of a dynamic wormhole}

  One method to detect a wormhole is by observing its shadow. When an observer looks at 
  a wormhole with a luminous background behind it, he/she sees a dark spot. It is the so-called shadow,
  which emerges due to the fact that some of the photons from the luminous background are captured 
  by the wormhole. We assume that light sources exist only in the observer’s Friedmann universe and 
  that there are no light sources inside the wormhole. We also do not consider photons arriving from 
  the other universe, as our small parameter $\eta$ approximation is not applicable in that case.

  In stationary scenarios, the boundary of a wormhole shadow is determined by photons moving along 
  cyclic orbits. However, in our dynamic case, such orbits do not exist, and the shadow boundary is
  determined by photons with the smallest angular momentum $L$ that reach the observer at the point
  $\chi_{\rm obs}$ at the observation time $\tau_{\rm obs}$.

\begin{figure}[h!]
\centering
\includegraphics[width=0.99\linewidth]{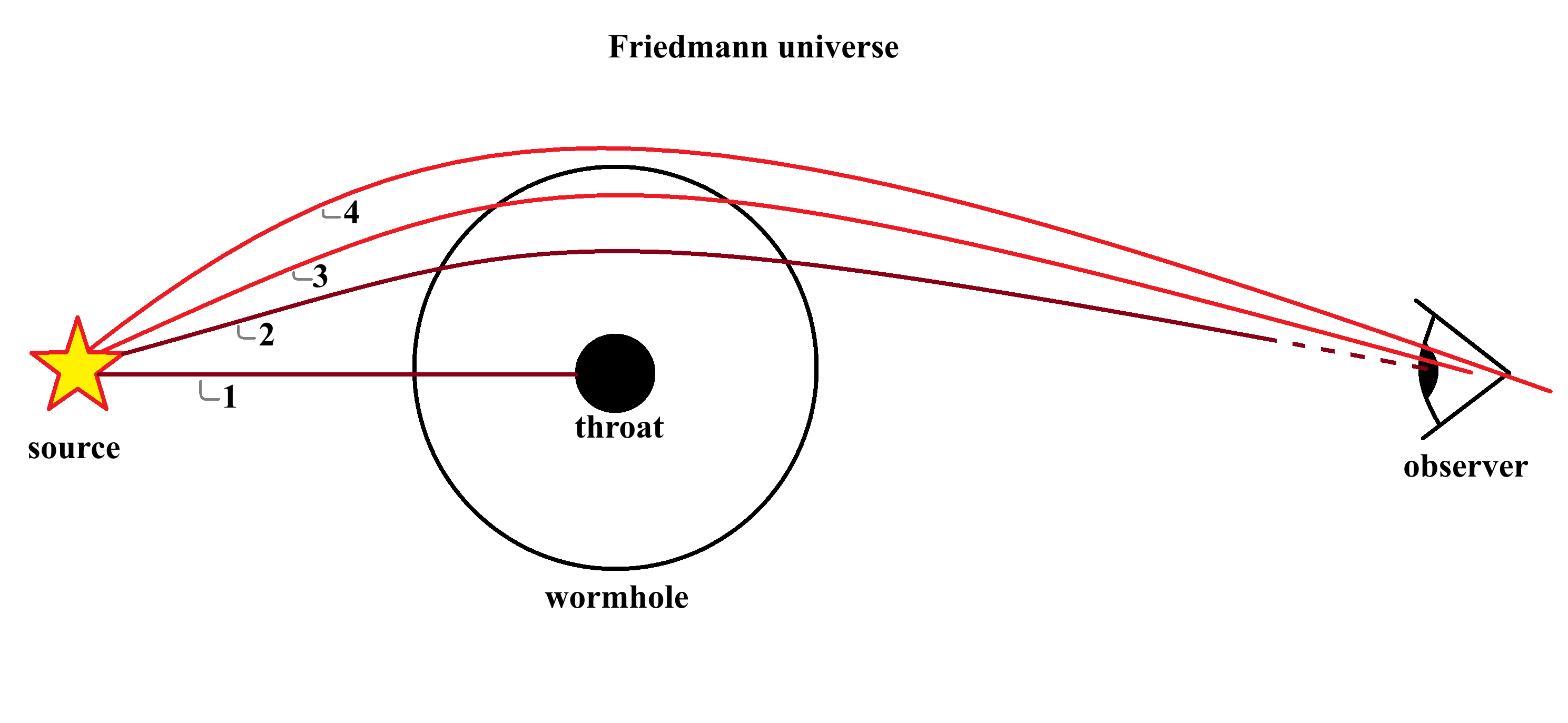}
\caption{\small
		Schematic representation of photon motion from a light source to an observer through a 
		wormhole. Photons 3 and 4 moving along the red trajectories have a higher angular momentum 
		$L$ and reach the observer by the time of observation. Photons 1 and 2, moving along the brown
		trajectories, have a lower angular momentum $L$ and do not reach the observer by the time 
		of observation. However, photon 2 does it later (brown dashed line).}
	\label{scheme}
\end{figure}

  Figure \ref{scheme} shows a schematic representation of photon motion from a light source to the
  observer through the wormhole. All photons are emitted simultaneously. Photons 3 and 4 (red trajectories)
  have higher angular momentum $L$ and reach the observer by the observation time. Photons 1 and 2
  (brown trajectories) have lower angular momenta $L$ and do not reach the observer by that time. 
  Thus, at this moment, the shadow boundary is formed by photon 3. However, after some time, 
  photon 2 will eventually reach the observer, and it will determine the shadow boundary, meaning that 
  the shadow size decreases. Hence, the observation time is the first factor affecting the size of the
  wormhole shadow.

\begin{figure}[h!]
\centering
\begin{minipage}[h]{0.7\linewidth}
		\includegraphics[width=1\linewidth]{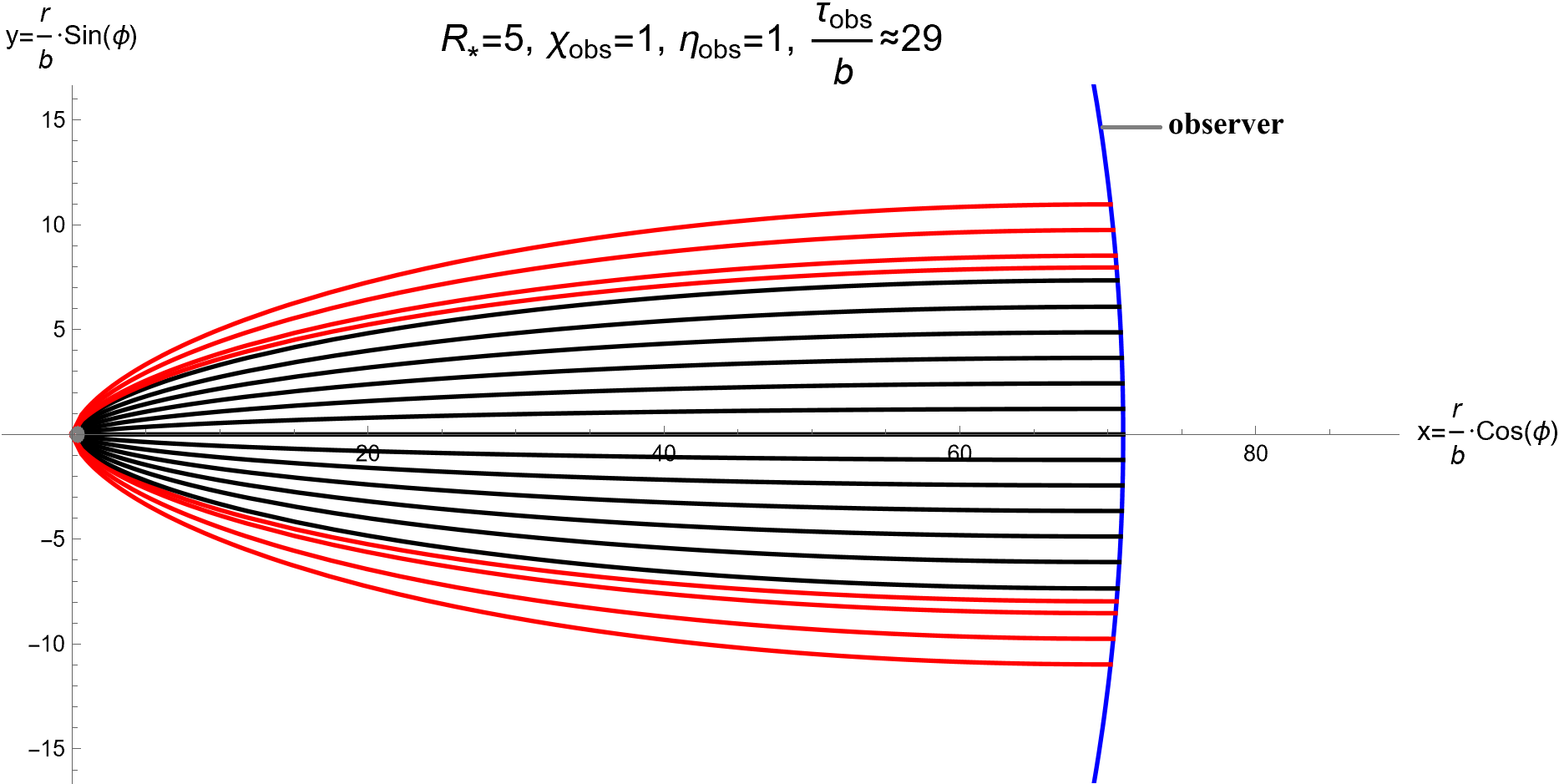}
\end{minipage}\\
	\vspace{3 ex}
\begin{minipage}[h]{0.7\linewidth}
		\includegraphics[width=1\linewidth]{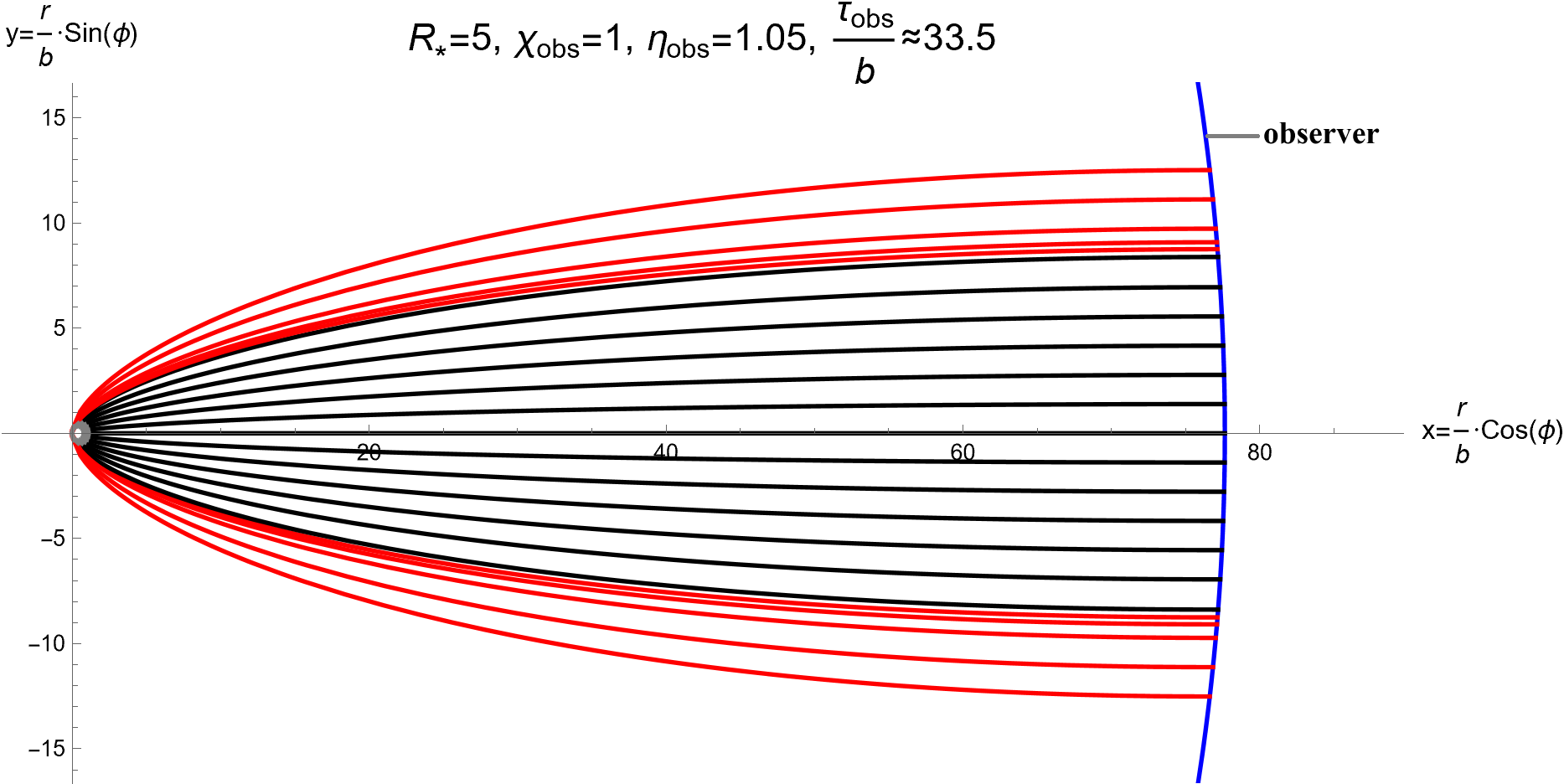}
\end{minipage}\\
	\vspace{3 ex}
\begin{minipage}[h]{0.7\linewidth}
		\includegraphics[width=1\linewidth]{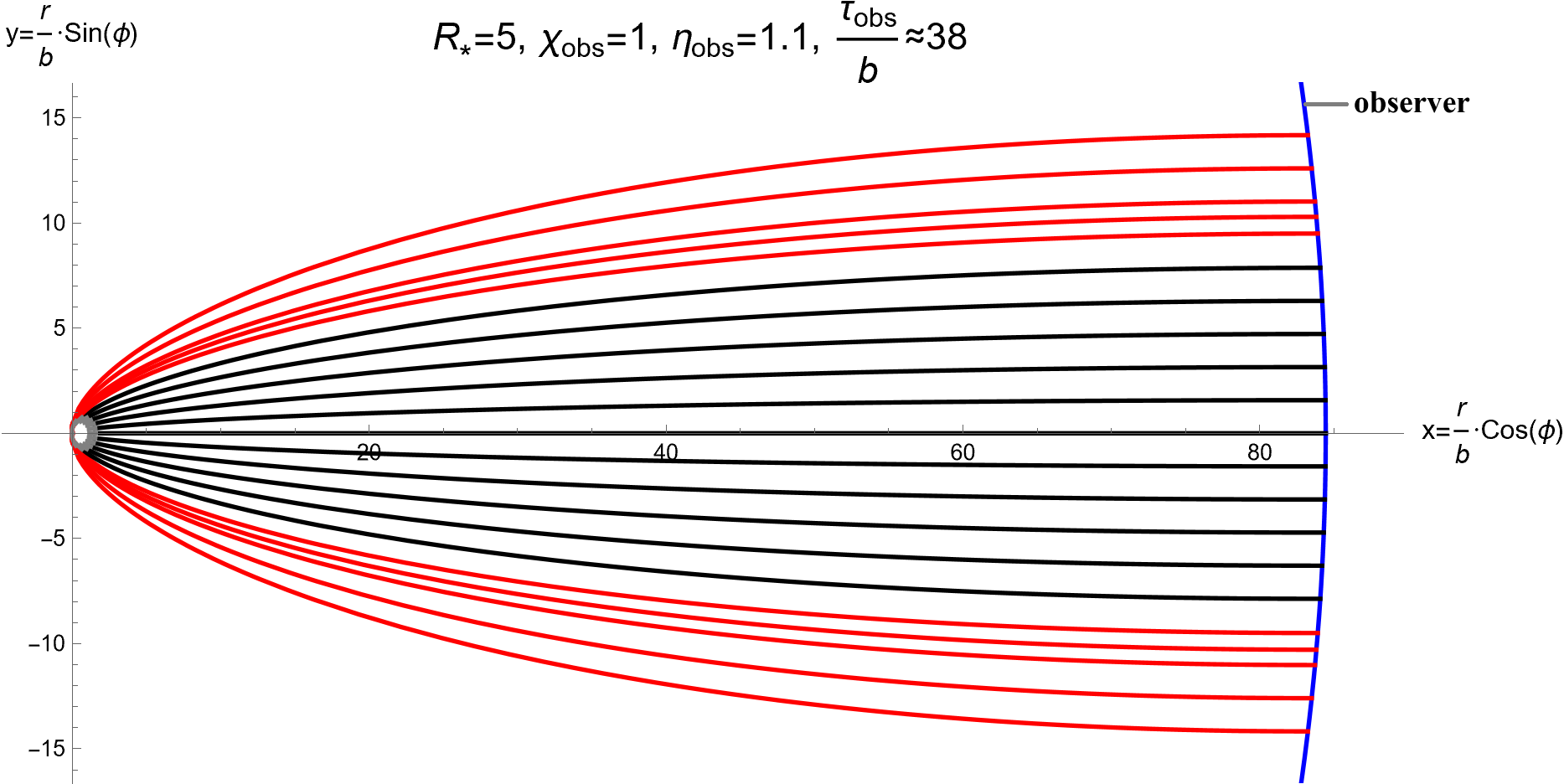}
\end{minipage}
\caption{\small
		Photon paths $r(\phi)$ in a Friedmann universe with a dynamic wormhole, 
		with $R_*=5$, at different observation times $\tau_{\rm obs}$.}
	\label{AllTraj}
\end{figure}

  The second factor affecting the shadow size is the expansion of the Friedmann universe. 
  Figure \ref{AllTraj} provides a numerical simulation of photon motion in a Friedmann universe 
  containing a dynamic wormhole. The images have been constructed for the case of photons that 
  began to move parallel to the $x$ axis from the point $\chi_{\rm obs}$ (blue line), $\tau_{\rm obs}$ 
  to the left, i.e., back in time. Some of the photons (black paths) fall into the wormhole and do not 
  have time to get out of it by $\tau = 0$. Another part of the photons (red paths) either move only in 
  the Friedmann universe or fall into the wormhole but manage to get out of it into the observer's
  universe. It should be noted that when photons are in the wormhole space-time, the value of the
  parameter $\eta$ does not exceed $0.1$, which is consistent with our small $\eta$ approximation.
  Thus, if we ``invert'' the paths and consider them from left to right, it turns out that the photons 
  begin their motion at the time $\tau\approx 0$ and fly to the right towards the observer. In this case, 
  the ``red'' photons reach the observer $\chi_{\rm obs}=1$ by the observation time $\tau_{\rm obs}$,
  while ``black'' ones do not reach the observer and form a wormhole shadow. It is easy to see that 
  the shadow in the lower image ($\tau_{\rm obs}/b\approx38$) is larger than in the upper one 
  ($\tau_{\rm obs}/b\approx 27$). This occurs due to the expansion of the Friedmann universe, 
  which makes the shadow increase in size.

  To calculate the size of the shadow seen by the observer, it is necessary to obtain the coordinates 
  of the light ray coming to the observer's sky. Assuming that space-time is flat near the observer 
  at any time, they can be found in the following way \cite{Vasquez}:
\beq      
		\alpha_i=-r_{\rm obs}^2\sin\theta_{\rm obs}\left.\frac{d\varphi}{dr}\right|_{r_{\rm obs}}, 
	\qquad 	\beta_i=r_{\rm obs}^2\left.\frac{d\theta}{dr}\right|_{r_{\rm obs}},      \label{skycoor}
\eeq      
  where $r_{\rm obs}$ and $\theta_{\rm obs}$ are the observer's coordinates.

  Since the space-times under consideration are spherically symmetric, the boundary of the shadow 
  will be a circle with radius $\alpha_{\rm sh}$, formed by the photons that reach the observer 
  and have the lowest angular momentum $L_{\rm sh}$. Substituting the geodesic equations
  (\ref{Eta})--(\ref{Phi}) into \eq (\ref{skycoor}) for $\alpha_i$ and assuming 
  $\theta_{\rm obs}=\pi/2$,  we obtain the shadow radius as
\beq      
	\alpha_{\rm sh}=\frac{a_0 L_{\rm sh}\left(1-\cos(\eta_{\rm obs})\right)^2}
	{\cos(\chi_{\rm obs})\left(1-\cos(\eta_{\rm obs})\right)
	\sqrt{K-\dfrac{L_{\rm sh}^2}{\sin^2(\chi_{\rm obs})}}+\sin(\chi_{\rm obs})
	\sin(\eta_{\rm obs})\sqrt{K}}\,.                        \label{radsh}
\eeq      
  The angular size of the shadow is given by
\beq      
		d_{\rm sh}=2\arctan\left(\frac{\alpha_{\rm sh}}{r_{\rm obs}}\right).            \label{sizesh}
\eeq      

\begin{figure}[h!]
\centering
\includegraphics[width=0.8\linewidth]{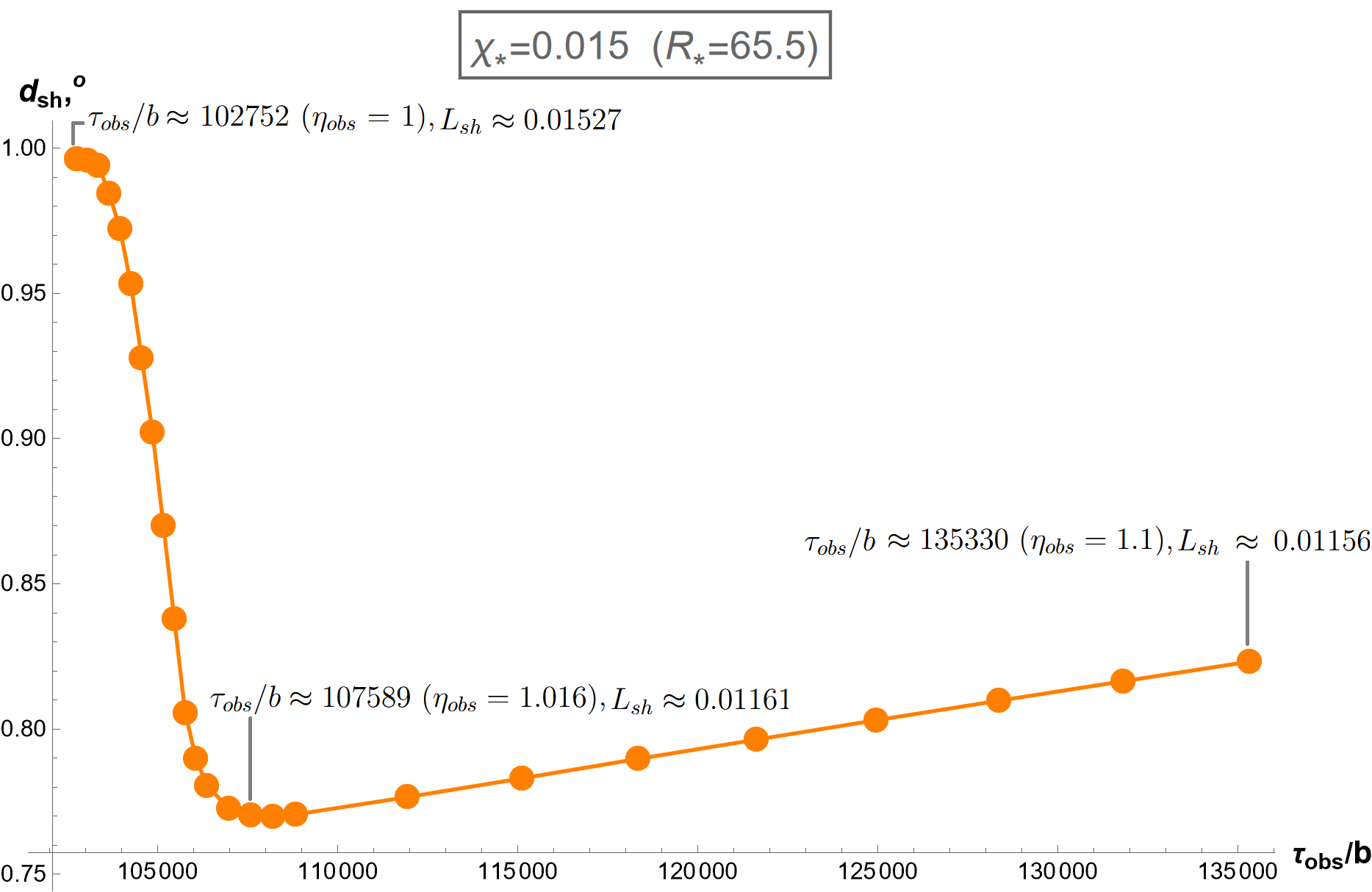}
\caption{\small
		The \wh\ shadow angular size $d_{\rm sh}$ for a \wh\ with 
		$\chi_*\approx 0.015(R_*\approx 65.5), \,k=0.1$ vs. observation time  $\tau_{\rm obs}/b$.}
	\label{shadIm}
\end{figure}

  Figure \ref{shadIm} shows the dependence of the angular size of the \wh\ shadow $d_{\rm sh}$ 
  for $\chi_*=0.015\, (R_*=65.5)$ on the observation time. As already mentioned, the size of the 
  shadow of a given wormhole is determined by both the parameter $L$ of the photons that reach 
  the observer and by the expansion of the universe itself. Therefore, the observation time 
  dependence of the shadow size looks somewhat unusual. 
  At the beginning of the observation, the shadow size decreases. This occurs because photons with 
  lower angular momentum $L_{\rm sh}$ gradually reach the observer, and this decrease is not
  immediately compensated by the expansion of the universe. However, at later stages of observation, 
  the angular momentum of the photons slowly decreases to the minimum value at which there is 
  a turning point, and the shadow size increases due to the expansion of the Friedmann universe.

  We will also obtain the dependence of the shadow size $d_{\rm sh}$ on the boundary value $\chi_*$.
  To do that, it is sufficient to find the minimum value $|L_{\min}|$ of photons having a turning 
  point for each wormhole, since precisely they form the shadow boundary at the observation time 
  $\eta_{\rm obs}=1.1$ if the observer is at the point $\chi_{\rm obs}=1$. From the expression 
  (\ref{Rturn}) we get
\beq      
		|L_{\min}|= \frac{3\,C}{5}\left(\frac{9\,b}{2}\right)^{1/3}.
\eeq      
  From the conditions  (\ref{match}) we obtain that the parameter $b$ is related to the \wh\ size   
  as $b=a_0(\sin\chi_*)^{3+2k}$. Since $\chi_* \ll 1$, $b\approx a_0\, \chi_*^{3+2k}$ and 
  $|L_{\min}|= (3\,C/5)\left(9\,a_0/2\right)^{1/3}\chi_*^{1+2k/3}$. From the equality of the values
  $d\tau/ds$ on the junction surface as calculated in the Friedmann universe and in the wormhole, 
  we can find that $\sqrt{K}\approx(3\,C/5)\left(9\,a_0/2\right)^{1/3}$ for all photons because 
  of the small $\eta$ approximation. We thus obtain
\beq      
		|L_{\min}|\approx \sqrt{K}\,\chi_*^{1+2k/3}.
\eeq      
  Now one should substitute this expression to \eq (\ref{radsh}). However, let us first consider the
  term with $L_{\rm sh}$ in the denominator. Replacing $L_{\rm sh}$ with $L_{\min}$, we obtain 
\[   
		\sqrt{K-\frac{L_{\min}^2}{\sin^2(\chi_{\rm obs})}}
		=\sqrt{K}\sqrt{1-\chi_*^{2+4k/3}\sin^{-2}(\chi_{\rm obs})}\approx\sqrt{K}. 
\]  
  Now the quantity $L_{\rm sh}$ remains only in the numerator, and in the approximation of 
  small angles we arrive at
\beq      
		d_{\rm sh}\sim \chi_*^{1+2k/3}.                      \label{powerdep}
\eeq 

\begin{figure}[h!]
\centering
\includegraphics[width=0.8\linewidth]{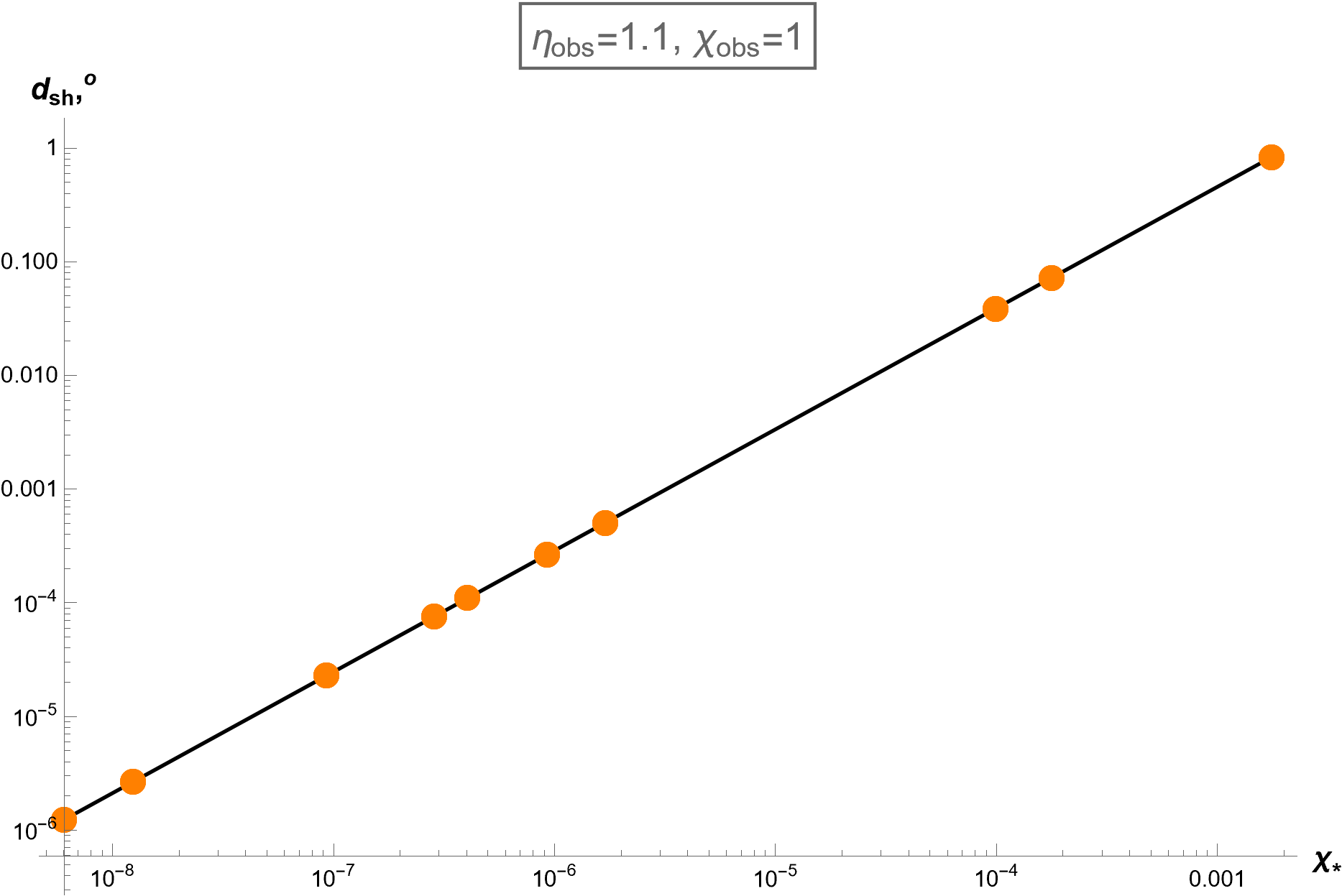}
\caption{\small
		The angular size of the wormhole shadow $d_{\rm sh}$ as a function of the \wh\ boundary 
		size $R_*$ at observation time $\eta_{\rm obs}=1.1$. Orange dots represent the values of 
		$\chi_*$ and $d_{\rm sh}$ taken from Table \ref{estimates}. The black line represents the 
		power-law dependence we have obtained. The observer is located at the point 
		$\chi_{\rm obs}=1$.}
	\label{shadowsIm}
\end{figure}

  Figure \ref{shadowsIm} shows the dependence of the angular size of the wormhole shadow 
  $d_{\rm sh}$ on the boundary value $\chi_*$ at the observation time $\eta_{\rm obs}=1.1$. 
  The observer is located at the point $\chi_{\rm obs}=1$. The orange dots represent the values 
  of $\chi_*$ and $d_{\rm sh}$ taken from Table\,\ref{estimates}. The black line represents the 
  power-law dependence we have obtained (\eq \ref{powerdep}).

\section{Conclusion \label{sec5}}

  In this paper, we have studied in detail the shadow of a dynamic traversable wormhole 
  inscribed into a closed dust-filled Friedmann universe.
  We have shown that the shadow 
  is determined by photons with the minimum angular momentum $L_{\rm sh}$ that reach a 
  particular distant observer by the observation time $\tau_{\rm obs}$. 

  Supposing $\eta\ll 1$, where $\eta$ is an auxiliary parameter used to express the solution 
  $r(R,\tau)$ in a parametric form, see \rf{rtn},  
  we derived an expression for the size of the wormhole shadow 
  and investigated the dependence of the shadow size on the observation time and on the boundary 
  value $\chi_*$ of the wormhole, which characterizes the size of the wormhole region. 

  We have obtained that the angular size of the shadow $d_{\rm sh}$ exhibits a non-monotonic
  dependence on the observation time. At early times, the shadow size decreases as photons with 
  smaller angular momentum gradually reach the observer. At later times, the expansion of the 
  Friedmann Universe becomes the dominant factor that leads to an increase in the shadow size.
  We have also derived a power-law dependence of the shadow size $d_{\rm sh}$ on the value 
  $\chi_*$ of the \wh\ region boundary: $d_{\rm sh}\sim \chi_*^{1+2k/3}$. This analytical result 
  was confirmed by our numerical calculations.

  It should be noted that these results are incomplete since we performed all calculations in the 
  small $\eta$ approximation and assumed $\tau_0(R) = 0$. That is, we considered only the 
  initial  stage of the wormhole evolution, for a \wh\ that appears simultaneously with the
   whole universe.  Also, due to the small $\eta$ approximation , we could not consider 
   photons that come to the observer from another universe since the parameter $\eta$
   significantly increases near the wormhole throat. Thus, in a future work it would be 
   interesting to abandon the small $\eta$ approximation and also to consider wormhole 
   models with different  $\tau_0(R)$, that is, \whs\ emerging at different cosmic times.
   
   In addition, as noticed in \cite{BroKasSus:2023}, the density of dust matter at the outskirts 
   of \wh\ regions is much smaller that the current mean density of the surrounding Friedmann
   universe, which allowed us to guess that such \wh\ regions could be related to the observed 
   voids in the distribution of matter in our Universe. Such a relationship could be relevant 
   even for \whs\ born together with the Universe: their rapidly evolving throats may have
   disappeared long ago while the outer volumes of \wh\ regions survive till nowadays. 
   This relationship can be one more promising subject of future studies.
   
\subsection*{Funding}

  KB was supported by the Ministry of Science and Higher Education of the Russian Federation, 
  Project "New Phenomena in Particle Physics and the Early Universe" FSWU-2023-0073.
  VA and SS were supported by the Russian Science Foundation grant No. 25-22-00163.


\end{document}